# DQSB: A Reliable Broadcast Protocol Based on Distributed Quasi-Synchronized Mechanism for Low Duty-Cycled Wireless Sensor Networks

Yun Wang[1] and Peizhong Shi[1] and Kai Li[1] and Jie Wu[2]

[1]School of Computer Science and Engineering, CNII, Southeast University, Nanjing, 211189, P.R. China
`{yunwang, peizhongshi, newlikai}@seu.edu.cn`
[2]Department of Computer and Information Sciences, Temple University, Philadelphia, PA 19122
`jiewu@temple.edu`

## ABSTRACT

*In duty-cycled wireless sensor networks, deployed sensor nodes are usually put to sleep for energy efficiency according to sleep scheduling approaches. Any sleep scheduling scheme with its supporting protocols ensures that data can always be routed from source to sink. In this paper, we investigate a problem of multi-hop broadcast and routing in random sleep scheduling scheme, and propose a novel protocol, called DQSB, by quasi-synchronization mechanism to achieve reliable broadcast and less latency routing. DQSB neither assumes time synchronization which requires all neighboring nodes wake up at the same time, nor assumes duty-cycled awareness which makes it difficult to use in asynchronous WSNs. Furthermore, the benefit of quasi-synchronized mechanism for broadcast from sink to other nodes is the less latency routing paths for reverse data collection to sink because of no or less sleep waiting time. Simulation results show that DQSB outperforms the existing protocols in broadcast times performance and keeps relative tolerant broadcast latency performance, even in the case of unreliable links. The proposed DQSB protocol, in this paper, can be recognized as a tradeoff between broadcast times and broadcast latency. We also explore the impact of parameters in the assumption and the approach to get proper values for supporting DQSB.*

## 1. INTRODUCTION

A Wireless Sensor Network (WSN) consists of a large number of small and low cost sensor nodes powered by small batteries and equipped with various sensing devices to observe events in the real world [1-4]. Usually, for many applications, once a WSN is deployed, probably in an inhospitable terrain, it is expected to gather required data for a certain period of time, which can reach a length of years. To bridge the gap between limited energy supplies and network lifetime, a WSN has to operate in a low duty-cycled manner, where nodes schedule themselves to be active for a brief period of time and then stay asleep for a long period of time [5, 6]. There are two types of duty-cycled WSNs, i.e. asynchronous sleep scheduling, where each sensor keeps a sleep schedule independent of another, and synchronous sleep scheduling, where sensors make synchronized periodic duty cycling with their neighboring nodes to support broadcast or unicast and reduce the idle listening energy cost. Any sleep scheduling scheme has to ensure that data can always be routed from source to sink [7].

Usually, sleep schedules are completely uncoordinated. Due to the variation of awake time and duration of the active interval, the whole network is more than often disconnected, and delay encountered in packet delivery due to loss in connectivity can become a critical problem. As a result, a path from source to sink may not always be available, and a sufficient number of nodes have to remain awake to ensure the existence of such a path. Consequently, data is stored at a node till its proper neighboring node wakes up and delivers the data to the sink. This approach would delay the delivery of messages to a sink considerably.

The existing works based on synchronization assume that there are usually multiple neighbors available at the same time to receive the multicast/flooding message sent by a sender. This is not true in low duty-cycled asynchronous WSNs. Furthermore, synchronization is another issue that is difficult to achieve, especially over multiple hops. Periodic synchronization messages may become costly. Usually, synchronization protocols are complex and difficult to implement in large scale WSNs. Without synchronized sleep scheduling, B-MAC [8], WiseMAC [9] and X-MAC [10] are based on asynchronous sleep intervals and proven to be energy-efficient in scenarios with low or varying traffic loads. Unfortunately, they cannot be directly applied to broadcast applications because of their design intentions for unicast.

Multi-hop broadcast is an important network service in WSNs, especially for applications such as code update, remote network configuration, route discovery, and so on. Distinguished from the broadcast problem in always-on networks, two additional features make multi-hop broadcast in low duty-cycled WSNs become a new challenging issue. Firstly, a node which broadcasts a message once cannot guarantee that the message is received by all of its neighboring nodes simultaneously, while this property is satisfied in an always-on network. To successfully broadcast a message, a sender has to transmit the same message more than once if other nodes do not wake up at the same time. Essentially, broadcasting in such a network is implemented by a number of unicasts. Secondly, in asynchronous duty-cycled WSNs, each node cannot be aware of its neighboring nodes' sleep schedules without neighboring discovery and information exchange protocols which require nodes to remain awake for enough time in order to aware their neighbors' sleep schedules.

Therefore, a question arises: Is it possible to maintain a high broadcast delivery rate and to exploit nodes' sleep schedules without the support of synchronization protocol at the same time, in asynchronous duty-cycled WSNs, where each sensor turns on and off independently and network connectivity is intermittent? Different from the existing related work, we propose a quasi-synchronization mechanism in order to coordinate nodes' duty-cycled behaviors in a distributed manner. It is quasi because nodes are not required to wake up at the exact same time. Sleep schedule adjustments stop if all the nodes except the sender are able to receive broadcast messages.

The main contributions of this paper are summarized as follows: (1) We propose a novel protocol DQSB by a mechanism of quasi-synchronization for multi-hop broadcast, which neither assumes time synchronization that requires all neighboring nodes wake up at the same time, nor assumes duty-cycled awareness that is difficult to use in asynchronous WSNs; (2) After broadcast process from a sink is finished under the quasi-synchronization mechanism, other nodes can build their paths to the sink for transmitting their sensed data after receiving the broadcast messages. Moreover, these paths exhibit less latency because of no or very little waiting time; (3) We develop a simulator based on the ONE simulator [11] and evaluate DQSB, including broadcast times and latency in different duty cycles, the impact of network size, reliability with unreliable links and less latency routing paths for reverse data collection from each node to broadcast source node, such as a sink. Simulation results show that the performance of DQSB satisfies the design goals.

The rest of the paper is organized as follows: Section 2 reviews the related work. Section 3 describes the models and assumptions of the solution to broadcasting and routing in duty-cycled WSNs. The design and implementation of DQSB are presented in Section 4. Simulation results are discussed and analyzed in Section 5, where the impact of parameters in the assumption and the approach to get proper values are explored for supporting DQSB. We conclude the paper in Section 6.

## 2. RELATED WORK

As we addressed in Section 1, multi-hop broadcast plays an important role in WSNs. Compared with the problem of broadcasting in always-on networks, neighbor connectivity becomes a more difficult problem in duty-cycled WSNs, where each node stays awake only for a fraction of time and neighboring nodes are not simultaneously awake for receiving data. A bunch of literature has addressed this problem.

According to the mechanism supported broadcasting, existing solutions are put into two categories, including synchronous and asynchronous sleep schedules. The former, such as S-MAC [12] and T-MAC [13], simplifies broadcast communication by letting neighboring nodes stay awake simultaneously. The latter solution has become increasingly attractive for data communication because of its energy efficiency. Due to space limitation, we focus only on reviews for broadcast solutions for asynchronous duty-cycled WSNs.

The protocols B-MAC [8], WiseMAC [9], and X-MAC [10] are based on asynchronous wake-up intervals and have proven to be more energy-efficient in scenarios with low or varying traffic load. B-MAC supports single-hop broadcasting in the same way for unicast, since the preamble transmission over an entire sleep period gives all of the transmitting nodes' neighbors a chance to detect the preamble and remain awake for data packets. B-MAC and WiseMAC broadcasting are each energetically costly and inefficient. When transmitting a frame, a full preamble is appended for alerting neighboring nodes to stay awake for the upcoming transmission of the broadcast frame. This broadcast approach with a full preamble wastes a lot of energy for sending and receiving, while the actual data transmission is often comparatively short. Without control measures for forwarder selection in multi-hop flooding, every broadcast message to be rebroadcast by every node will experience the wireless-channel characteristic broadcast storm problem. Consequently, the broadcast success ratio and latency performance decreases. X-MAC, a low power MAC protocol, substantially improves B-MAC's excess latency at each hop and reduces energy usage at both the transmitter and receiver by employing a shortened preamble approach. But broadcast support is not clearly discussed in that paper. X-MAC is not promising for broadcasting since the transmitter has to continually trigger the neighbors to wake up, no matter whether it has received or not.

The (k)-Best-Instants broadcast algorithm [14], calculating the best instants and transmitting the frame with a minimized preamble, can be more efficient than using a costly full-cycle preamble like WiseMAC. Its assumption is the sender is aware of their neighbors' individual schedules. Wang et al. [15] transformed the problem into a shortest-path problem with the same assumption of duty-cycle awareness, which makes it difficult to use it in asynchronous WSNs since duty-cycle awareness needs periodic time-synchronization due to clock drifting. Focusing on energy-harvesting networks, Gu et al. [16] introduce the proactive generic delay maintenance algorithm to minimize the amount of energy while satisfying an end-to-end delay bound specified by application requirements for sink-to-many communications in energy-harvesting networks. But nodes in the network must share their duty-cycled working schedules with neighboring nodes for the assumption of duty-cycle awareness, so as to know when they can send a packet to their neighbors with the support of local synchronization techniques [17].

Opportunistic routing and data forwarding in low duty-cycled networks have acquired a lot of attention in recent years [18, 19]. But none of these solutions investigates the broadcasting. ADB [20] and opportunistic flooding [21] were designed with a gossiping approach as long as the network is connected. ADB avoids the problems with B-MAC and X-MAC by dynamically optimizing the broadcasting at the level of transmission, to each individual neighboring node. It allows a node to go to sleep immediately when no more neighbors need to be reached and does not occupy the medium for a long time, in order to minimize latency before forwarding a broadcast. The effort in delivering a broadcast packet to a neighbor is adjusted based on link quality, rather than transmitting throughout a duty cycle or waiting through a duty cycle for neighbors to wake up. Basically, ADB belongs to the unicast replacement approach and it needs significant modification to existing MAC protocols for supporting broadcast. In [21], a design of opportunistic flooding has been proposed for low duty-cycled networks with unreliable wireless links and predetermined working schedules. It provides probabilistic forwarding decisions at a sender based on the delay distribution of next-hop nodes and a forwarder selection method to alleviate the hidden terminal problem and a link-quality-based backoff method to resolve. However, these protocols for duty-cycled WSNs, belonging to the unicast replacement approach for supporting broadcasting, mostly focus on unicast communication and cannot well support broadcasting since one-hop broadcasting in such cases means to deliver data multiple times to all neighbors, which may lack efficiency in large scale networks, and also lack energy efficiency in delivering large chunks of data for broadcasting.

Hybrid-cast [22, 23], with low latency and reduced message count, overcomes the disadvantages of replacement via pure unicast. Under Hybrid-cast, nodes must switch their wake-up schedule to stay awake for enough time slots for neighbor discovery and information exchange. Then, the online forwarder selection algorithm works and helps to reduce the broadcast count or redundant transmission for multi-hop broadcast.

In conclusion, the above protocols either prevent themselves from being widely used in realistic environments due to their assumptions, including duty-cycled awareness and neighbor discovery supporting in asynchronous duty-cycled WSNs, or only belong to the unicast replacement approach for supporting broadcasting. We focus on exploiting nodes' sleep schedules and make adjustment strategies to solve the multi-hop broadcast problem by a distributed and quasi-synchronized manner. Meanwhile, a broadcasting node, such as sink, is in charge of data collection broadcast periodically. Receiving the broadcast messages, other nodes can build their paths to the sink for transmitting their sensed data, where these paths have less latency since the advantage of our quasi-synchronization mechanism. Unlike B-MAC, WiseMAC, and X-MAC, a unicast message in our paper can be transmitted along a path learned from the quasi-synchronized broadcasting and eliminates the waiting time for both transmitter and receiver.

## 3. PROBLEM DESCRIPTION

### 3.1. System Model

Suppose that in duty-cycled WSNs, there are $p$ sensor nodes $N = \{n_1, n_2, \cdots, n_p\}$, $|N| = p$, working in two states: active and sleep states. All nodes have their own sleep schedules and are able to adjust their sleep schedules if necessary. A network is denoted by a time-dependent graph $G(t) = (N, E(t))$, where $N$ is a complete set of nodes in the network and $E(t)$ is a set of undirected edges at time $t$. Evolving graph [24] is used to capture the dynamic characteristics, especially node intermittent connectivity with its neighbors. An evolving graph $G(t) = N(N, E(t))$ is connected during $t$, where $t \in [0, T]$ and $T$ is one cycle length, if no isolated edge and vertex exists in $G(t)$. Let $L_{i,j}(t_{i,j}, T_{i,j})$ denote the intermittent connective link between nodes $n_i$ and $n_j$, $n_i, n_j \in N$ and $L_{i,j}(t_{i,j}, T_{i,j}) \in E(t)$. The link begins at $t_{i,j}$ and keeps a period of time $T_{i,j}$. For its bidirectional property, $L_{i,j}(t_{i,j}, T_{i,j}) = L_{j,i}(t_{j,i}, T_{j,i})$ is established and $L_{i,j}(t_{i,j}, T_{i,j}) \in E(t)$. If $n_i$ wakes up earlier than $n_j$, i.e. $t_i^a < t_j^a$, then $t_{i,j} = t_j^a$. Otherwise, $t_{i,j} = t_i^a$. All the variables used in the paper are listed in Table 1.

Table 1. Variables and their significance.

| Variable | Description |
|---|---|
| $t_i^a$ | Node $n_i$'s time to wake up |
| $t_i^s$ | Node $n_i$'s time to sleep |
| $T_a$ | Node $n_i$'s active duration time |
| $T_s$ | Node $n_i$'s sleep duration time |
| $T_0$ | The time for transmitting a broadcast message |
| $R$ | Node's Communication radius |
| $\rho$ | Network density |
| $N_R$ | Number of nodes in circle area $\pi R^2$, namely $N_R = \pi R^2 \cdot \rho$ |
| $\lambda$ | Parameter of Poisson distribution for intermittent connectivity |
| $\lambda_0$ | $\lambda_0 = N_R / T_a$ |
| $\alpha$ | Parameter for selecting $\lambda$ that $\lambda = \alpha \cdot \lambda_0$ for intermittent connectivity |
| $P_\lambda$ | Broadcast success ratio under parameter $\lambda$ |
| $t_i^{send}$ | Node $n_i$'s time to forward broadcast |
| $T_i^{backoff}$ | Node $n_i$'s exponential backoff time |
| $M_i$ | Node $n_i$'s message buffer |
| $B_i$ | Node $n_i$'s beacon message |
| $footer$ | Transmitting with DATA to reduce broadcast redundancy |
| $B_i.Id$ | The max Id of the broadcast message received by node $n_i$ |
| $\Delta T(i)$ | The latency of the broadcast message received by node $n_i$ |
| $\Delta T^k(i)$ | Node $n_i$'s latency $\Delta T(i)$ in case k |
| $t_i^{a,k}$ | Node $n_i$'s wake-up time $t_i^a$ in case k |

There are two short packets, i.e., *beacon* and *footer*, used in DQSB, as shown in Figure 1. A *beacon* is used by a node to announce its active state when it wakes up. It includes *Id*, *Node_id* and *Wakeup_time*. Field *Id* is used to help its neighbor make forwarding decisions or trigger its neighbors to adjust their sleep scheduling to receive its broadcast message. So, in *beacon*, the value of *Id* will be set to the maximum sequence number of the received broadcast message in its message buffer and it will be updated dynamically. Field *Node_id* and *Wakeup_time* are used to tell neighboring nodes when the node wakes up. The *footer* indicating the transmitting for its neighboring nodes contains fields such as *Forwarder*, *Receivers*, *Message_id* and *End_time*. Suppose that forwarder $n_i$ and $n_j$ will transmit the broadcast message with the same *Message_id* at time $t_i^{send}$ and $t_j^{send}$ respectively ($t_i^{send} < t_j^{send}$). However, if $n_j$'s receivers are contained in $n_i$'s and $n_j$ can hear the *footer* from $n_i$, then $n_j$ will abort this forwarding task at time $t_j^{send}$ in order to reduce broadcast redundancy. Otherwise, $n_j$ must start its forwarding task with backoff mechanism to avoid a collision when it learns from *End_time* ( $End\_time = t_i^{send} + T_0$ ) in $n_i$'s *footer* that $t_i^{send} < t_j^{send} < End\_time$, indicating $n_j$'s forwarding will take place during $n_i$'s transmitting. Besides, *footer* and *DATA* carried in a packet will be transmitted as a broadcast message in DQSB.

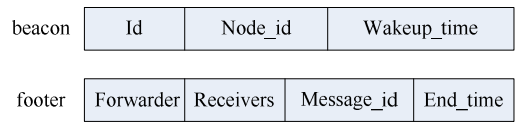

Figure 1. Packet structures of *beacon* and *footer*.

An assumption used in [25] as well is that the time interval between two asynchronous duty-cycled nodes is an exponentially distributed random variable with average $T_a/N_R$, where $T_a$ is the length of the node's active period after waking up and $N_R$ is the number of nodes in the circle area of communication radius $R$. A sequence of waking up behavior of nodes is represented by a Poisson process. The probability $P_t$ that more than one node wakes up in a period $t$ is formulated as follows:

$$P_t = 1 - e^{-\frac{N_R}{T_a} \cdot t} - \frac{N_R}{T_a} \cdot t \cdot e^{-\frac{N_R}{T_a} \cdot t} = 1 - (1 + \frac{N_R}{T_a} \cdot t) \cdot e^{-\frac{N_R}{T_a} \cdot t}$$

From the above equation, if the period $t$ is extremely short (approaching zero), $P_t$ approaches zero. Hence, in an asynchronous duty-cycled WSN, the probability that more than one node wakes up simultaneously is almost zero. Consequently, this helps us to ignore the collisions happening among nodes when they wake up and immediately send a short packet. With these assumptions, in asynchronous duty-cycled WSNs, we assume that there is a $\lambda$ ($\lambda = \alpha \cdot \lambda_0$, $\lambda_0 = N_R/T_a$) such that for all $\lambda'$ ($\lambda' \geq \lambda$), an evolving graph G(t) is intermittent connectivity during time t, where t $\in$ [0,T] and T is one cycle length that $T = T_a + T_s$. For properly selecting $\lambda$ to ensure the assumption, we will explore the impact of $\lambda$ and get the smallest value of $\alpha$ in Section 5.

### 3.2. Conditions for Successful Broadcast

In an asynchronous duty-cycled WSN, suppose the broadcast time of node $n_i$ is set to $t_i^{send}$ if it is a broadcast forwarder. There is $t_i^{send} = t_i^s - T_0$. $n_j$ receives a message sent by $n_i$ if and only if the duration $T_{i,j}$ for the link $L_{i,j}(t_{i,j}, T_{i,j})$ satisfies the following conditions. As shown in Figure 2, there are two conditions:

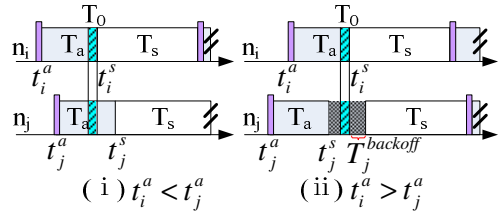

Figure 2. Two conditions of timing relationship between transmitter $n_i$ and its receiver $n_j$.

(1) If $t_i^a < t_j^a$, link $L_{i,j}$'s duration time $T_{i,j}$ satisfies $T_{i,j} \geq T_0$. In this case, the transmitter $n_i$ wakes up earlier than its receiver $n_j$. If the link $L_{i,j}(t_{i,j}, T_{i,j})$ between nodes $n_i$ and $n_j$ satisfies the condition, then $t_i^s - t_j^a \geq T_0$, where equation $t_i^s - t_j^a = T_{i,j}$ holds. So, the condition is converted to $T_{i,j} \geq T_0$.

(2) If $t_i^a > t_j^a$, $n_j$ adjusts its sleep schedule by $t_j^s = t_i^s + T_j^{backoff}$ if it cannot receive broadcast messages in an opportunistic way from other nodes. This case indicates that if $n_i$, serving as the transmitter, wakes up later than the receiver $n_j$, $n_j$ cannot receive $n_i$'s broadcast message since $n_j$ has already switched to sleep state. Therefore, $n_j$ must prolong its active time and move to sleep state later than the transmitter $n_i$. It is expressed as $t_j^s = t_j^s + (t_i^a - t_j^a + T_j^{backoff})$, which is further converted to $t_j^s = t_i^s + T_j^{backoff}$ due to $t_j^s - t_j^a = T_a$ and $t_i^a + T_a = t_i^s$. Then, this case is converted to the condition 1.

In summary, all the sleep schedules of nodes need to satisfy the condition 1 in order to let receivers properly receive broadcast messages.

### 3.2. Quasi-Synchronization Mechanism

A broadcast protocol aims to generate a broadcast tree from the broadcast source node to all the other nodes, and the sleep scheduling relationship between any node and its parent node satisfies the condition that a sender wakes up earlier than its receiver. Quasi-synchronization mechanism proposed in this paper is responsible for this undertaking goal. If a receiver wakes up earlier than its parent node, it is called timing inversion. The mechanism firstly helps nodes determine whether there is a timing inversion in its path of the broadcast tree. If it is, the mechanism requests the related nodes to adjust their sleep schedules. Consequently, the condition remains true among all the nodes in the broadcast tree. All the nodes reach a quasi-synchronized state. Quasi, here, means that all the nodes may not wake up at the same time, but they are able to receive all the broadcast messages sent by the root of the broadcast tree.

The quasi-synchronization mechanism needs to handle 4 cases due to the relationship between a sender and its direct receiver. Suppose that $n_i$ transmits a broadcast message at time $t_i^{send}$, $t_i^{send} = t_i^s - T_0$ as shown in Figure 3.

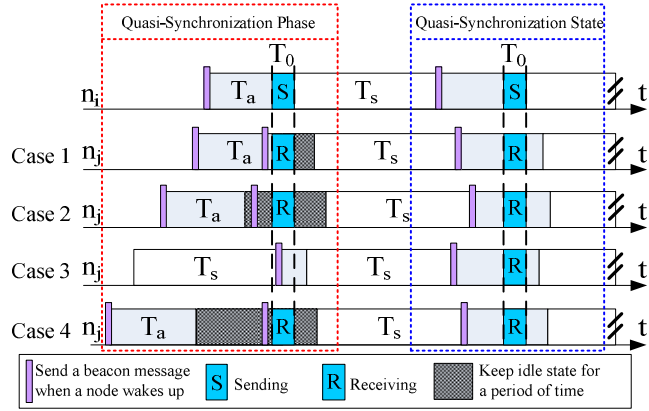

Figure 3. Suppose $n_i$ is a broadcast forwarding node and $n_j$ is its receiver. Different cases are considered in quasi-synchronization mechanism and their adjustment strategies: (Case 1 and Case 2) Early Sleep. (Case 3) Late Wake-up. (Case 4) Isolated Node.

(1) $n_j$ wakes up before $n_i$. $n_j$ fails to receive $n_i$'s broadcast message. If $n_j$ does not receive the broadcast message from other neighbor nodes during $[t_{n_j,a}, t_{n_i,send} - T_j^{backoff}]$ (case 1) or $[t_{n_j,a}, t_{n_j,s}]$ (case 2), $n_j$ retransmits its *beacon* at time $t$ ($t = t_i^{send} - T_j^{backoff}$) to trigger $n_i$'s forwarding decision. Both case 1 and case 2 are due to receivers' early sleep.

(2) $n_j$'s wake-up happens during $n_i$'s transmitting time. $n_j$ cannot receive $n_i$'s broadcast message correctly. This case is called late wake-up problem and its corresponding solution for $n_j$ is to adjust its next sleep time and sets its $t_{j,s}$ to $t_{j,s} - T_j^{backoff}$.

(3) Case 4 is due to the isolated node problem. When $n_j$ finds that it cannot receive anything else during its active period of $[t_j^a, t_j^s - T_0]$, it prolongs its active time until it receives its neighboring nodes' wake-up message, such as *beacons*. If $n_j$ can learn whether any of their neighbors have the broadcast message it wants from the receiving *beacons*, $n_j$ can decide whether or not to retransmit its *beacon* at time $t$ ($t = t_{n_j,a} + T_j^{backoff}$) to trigger $n_i$'s forwarding and wait for the the upcoming broadcast message. Then, $n_j$ switches to sleep state at $t_{j,s}$ ($t_{j,s} = t_{i,s} + T_j^{backoff}$).

After the adjustments of nodes' sleep schedules in quasi-synchronization in Figure 3, the timing inversion problem is solved.

## 4. DQSB PROTOCOL DESIGN

### 4.1. Overview

DQSB aims at giving a solution to multi-hop broadcast and routing in asynchronous duty-cycled WSNs without time synchronization and duty-cycled awareness, and helps nodes to forward broadcast messages and transmit sensed data to a broadcast source node, to reduce broadcast redundancy and keep relative tolerant broadcast latency performance. It is regarded as a joint design approach to achieve reliable broadcast and less latency routing paths for reverse data collection to a broadcast source node. The main idea of the protocol is to let nodes locally develop their own views of the sleep schedule of the whole network. Although in a system viewpoint, the nodes' sleep schedules are not strictly synchronized, the adjustments of nodes' sleep schedules are good enough to guarantee that all the nodes receive broadcast messages. Under DQSB, a data collection node, such as sink, periodically broadcasts messages to inform other nodes to transmit their sensing data to it. The other nodes firstly try to receive broadcast messages in an opportunistic way. If this fails, they will trigger one of their neighbor nodes' forwarding decisions and adjust their duty-cycle in a local and distributed manner for receiving. As a result, nodes' sleep schedules are coordinated. Every node learns its next hop to the sink by a minimal latency path due to no or less waiting time and no synchronization delay.

### 4.2. Distributed Quasi-Synchronized Broadcast

DQSB is composed of two basic components: (1) Forwarding decision, which helps nodes to know whether or not to forward received broadcast messages; and (2) Adaptive sleep scheduling adjustment, which is triggered when a node is aware of the upcoming transmissions which it misses. The node will adjust its sleep schedule in order to receive necessary messages. DQSB has five tasks to complete: forwarding decision, sending task abort, managing early sleep, tackling late wake-up and dealing with isolated nodes. The protocol is presented in Algorithm 1, where variables at each node have been described in Table 1. For consistency with Figure 2 and Figure 3, let's illustrate how $n_j$ finishes the following tasks with one of its neighboring nodes, such as $n_i$.

When a node $n_j$ wakes up, at first it drops expired messages and gets the newly received broadcast messages in its message buffer (line 1-2). Then, it transmits its *beacon* as a one-hop broadcast immediately, indicating the maximum identification of received broadcast messages. If the message buffer is empty, the value of *Id* in the *beacon* is set to -1 (line 3-5). After transmitting its *beacon* $B_j$, it will receive *beacons* from its neighboring nodes during the interval of $[t_j^a, t_j^s - T_0]$, such as $B_i$ from neighboring $n_i$ .(line 6-30).

***Task 1:*** *Forwarding Decision (line 7-10).* Forwarding decision is made by $n_j$ according to the received *beacons* from neighboring nodes during $n_j$'s active interval of $[t_j^a, t_j^s - T_0]$. $n_j$'s sending is triggered if and only if the condition $(M_j != null) \&\& (\exists m \in M_j, m.Id > B_i.Id, B_i \in beacons)$ is satisfied. If $n_j$ really forwards a message, its sending time $t_j^{send}$ is set to $t_j^s - T_0$.

***Task 2:*** *Sending Task Abort (line 22-24).* Sending task abort happens if and only if node $n_j$ listens to the channel and receives the *footer* before $t_j^{send}$, and the *footer* indicating the transmitting for its neighbor nodes at time $t_j^{send}$ has been done.

***Task 3:*** *Managing Early Sleep (line 11-19).* Early sleep occurs if and only if $n_j$ receives more than one *beacon* during its active period of $[(t_j^a, t_j^s - T_0)]$ and meets one of the following conditions: (1) $\forall m \in M_j, m.Id < B_i.Id, B_i \in beacons$ or (2) $M_j == null \&\& B_i.Id \neq -1, B_i \in beacons$. According to the analysis of Figure 2, the relationship between transmitter $n_i$ and its receiver $n_j$ satisfies the condition $t_i^a > t_j^a$. Case 1 and 2 are recognized

International Journal of Wireless & Mobile Networks (IJWMN) Vol. 4, No. 3, June 2012

as the early sleep problems discussed in Figure 3. Their solutions to trigger neighbors' transmitting are given, respectively: (1) If $(T_a - T_0 < T_{i,j} < T_a)$, $n_j$ updates its $B_j.Id$ when receiving broadcast messages during $[t_j^a, t_i^{send} - T_j^{backoff}]$ (Case 1); (2) If $(0 < T_{i,j} < T_a - T_0)$, $n_j$ updates its $B_j.Id$ when receiving broadcast messages during $[t_j^a, t_j^s]$ (Case 2). After updating its $B_j.Id$ during these times, if the condition $B_j.Id < B_i.Id$ is still satisfied, $n_j$ retransmits its beacon $B_j$ at time $t = t_i^{send} - T_j^{backoff}$ to trigger $n_i$'s forwarding. Then, $n_j$ only waits for the upcoming broadcast message. After receiving, $n_j$ adjusts its $t_j^s$ ($t_j^s = t_i^s + T_j^{backoff}$), namely condition $t_i^a < t_j^a$ holds in the following duty cycle.

---

**Algorithm 1: Distributed quasi-synchronized broadcast**

```
begin
  if(node n_j's wakeup time t_j^a arrive) then
    Move to idle state;
    M_j ← getMessageBuffer();
    if( M_j != null ) then B_j.Id ← M_j.getMaxId();
    else B_j.Id ← −1;
    Transmit its beacon B_j;
    While (t_j^a ≤ t ≤ t_j^s − T_0) do
      if(receive beacon B_i from n_i) then
        if( (M_j != null) & (∃m ∈ M_j, m.Id > B_i.Id) ) then
          t_j^send ← t_j^s − T_0;
          Wait t_j^send arrive and broadcast m;
        else if( ∀m ∈ M_j, m.Id < B_i.Id ) || (M_j == null && B_i.Id ≠ −1) ) then
          if( T_a − T_0 < T_{i,j} < T_a ) then
            Update B_j.Id when receives broadcast message during t_j^a < t < t_i^send − T_j^backoff;
          else if( 0 < T_{i,j} < T_a − T_0 ) then
            Update B_j.Id when receives broadcast message during t_j^a < t < t_j^s;
          if( B_j.Id < B_i.Id ) then
            t_j^s ← t_i^s + T_j^backoff;
            Retransmits its beacon B_j at time t (t = t_i^send − T_j^backoff) to trigger n_i's forwarding;
            Wait and receive broadcast message;
      else if(neighbor n_i is transmitting && t_i^send < t_j^a < t_i^s) then
        t_j^s ← t_j^s − T_j^backoff;
      else if( n_j receives footer before t_i^send ) then
        if(The footer indicates the transmitting for its neighbor nodes has been done) then
          Abort its sending at t_j^send;
      else if( n_j can't receive any message) then
        Prolong its active time until it receives neighbor nodes' beacons;
        if( B_i.Id > B_j.Id, B_i.Id ∈ beacons ) then
          Retransmits its beacon B_j at time t (t = t_j^a + T_j^backoff) to trigger n_i's forwarding;
          Wait and receive broadcast message;
          t_j^s ← t_j^s + T_j^backoff;
  if(node j's sleep time t_j^s arrive) then
    Move to sleep state
end
```

---

***Task 4:*** *Tackling Late Wake-up (line 20-21).* As case 3 shows in Figure 3, a late wake-up occurs if and only if $n_j$ listens to the channel and overhears neighboring node $n_i$'s transmitting during $[(t_j^a, t_j^s − T_0)]$ and $n_j$'s wake-up time $t_j^a$ satisfies condition $t_i^a < t_i^{send} < t_j^a < t_i^s$. However, here, $n_i$ is the transmitter, and the condition $T_{i,j} < T_0$ cannot satisfy the condition $T_{i,j} \geq T_0$ required by Condition (1) in Figure 2. The corresponding solution for $n_j$ is to set its $t_{j,s}$ to $t_{j,s} = t_{j,s} − T_{j}^{backoff}$.

***Task 5:*** *Dealing With Isolated Nodes (line 25-30).* Isolated nodes occur if and only if $n_j$ receives no *beacon* and listens nothing else during its active period of $[t_j^a, t_j^s - T_0)]$. Although this problem may not happen at the initial time of a WSN due to our assumption, it may really occur if nodes recover from failure or join in the network afterwards. As case 4 shows in Figure 3, when $n_j$ finds that it cannot receive any *beacon* message and anything else during its active period of $[t_j^a, t_j^s - T_0]$, it prolongs active time until it receives neighboring nodes' *beacons* e.g. $B_i$. If the condition that $B_i.Id > B_j.Id, B_i.Id \in beacons$ holds, $n_j$ retransmits its beacon $B_j$ at time $t$ ($t = t_{i,a} + T_i^{backoff}$) to trigger $n_i$'s forwarding and waits for the upcoming broadcast message. $n_j$ switches to sleep state at $t_{j,s}$ ($t_{j,s} = t_{i,s} + T_j^{backoff}$).

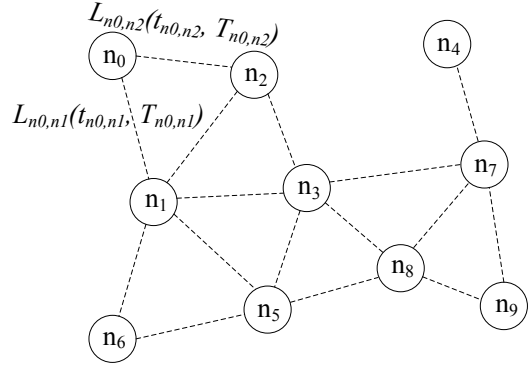

Figure 4. An evolving graph $G(t)$ of ten nodes in an asynchronous duty-cycled WSN. $G(t) = (N, E(t))$, where $N = \{n_0, n_1, \cdots, n_9\}$ and the dotted lines denote intermittent connective links $L_{i,j}(t_{i,j}, T_{i,j})$ between $n_i$ and $n_j$ with $L_{i,j}(t_{i,j}, T_{i,j}) \in E(t)$.

We further discuss DQSB in detail with an example. As shown in Figure 4, suppose there are ten nodes in an asynchronous duty-cycled WSN. $n_0$ is in charge of the data collection and periodically broadcasts some DATA packets to the other nodes. Figure 5 gives an overview of the operation sequences of DQSB regarding the scenario in Figure 4. In the forwarding decision phase, because $n_0$ is the broadcast source at first, $B_0.Id$ is set to 0 for the first broadcast message and is transmitted when it wakes up (Sequence 1). As $n_0$'s neighbors, both of $n_1$ and $n_2$ wake up before $n_0$, they can receive $n_0$'s beacon message $B_0$ and fail to receive $n_i$'s broadcast message because of early sleep problem. But when they can receive the broadcast message from other neighboring nodes, they will update their $Id$ values of the beacons during $[t_1^a, t_1^s]$ and $[t_2^a, t_0^{send} - T_2^{backoff}]$ respectively. Otherwise, $M_1 == null$, $M_2 == null$ and $B_0.Id \neq -1, B_0 \in beacons$ satisfy the condition (2) in Task3. Here, both $B_1.Id$ and $B_2.Id$ will be set to -1. Then they retransmit their *beacons* before $t_0^{send}$ (Sequence 2 and 3). When $n_0$ receives $B_1$ and $B_2$ from $n_1$ and $n_2$ respectively, any of them will trigger $n_0$'s sending at $t_0^{send}$ due to $B_0.Id > B_1.Id$ or $B_0.Id > B_2.Id$ (Sequence 4). When $n_1$ and $n_2$ receive the broadcast messages, they move to sleep state at time $t_1^s$ ($t_1^s = t_0^s + T_1^{backoff}$) and $t_2^s$ ($t_2^s = t_0^s + T_2^{backoff}$) respectively (Sequence 5 and 6). Suppose $n_3$ finds it cannot receive any *beacon* from neighboring nodes during its active period of $[t_3^a, t_3^s - T_0]$, it prolongs its active time till receiving neighboring nodes' *beacons*, e.g., $n_1$ and $n_2$. Then $n_3$ retransmits $B_3$ because of $B_1.Id > B_3.Id$ or $B_2.Id > B_3.Id$ (Sequence 7). After receiving $n_3$'s beacon $B_3$, both $n_1$ and $n_2$ will make a forwarding decision (Sequence 8 and 9). But in asynchronous duty-cycled WSNs, $n_1$ and $n_2$ do not transmit simultaneously according to their expected sending time. Since $n_2$ listens $n_1$'s transmitting and learns from the *footer* indicating the broadcast message $n_3$ wants has been transmitted by $n_1$, $n_2$ will abort the sending task at time $t_2^{send}$ to avoid collision and reduce broadcast redundancy (Sequence 10). Unfortunately $n_6$'s wake-up happens during $n_1$'s transmitting time, $n_6$ cannot receive the broadcast message correctly. Then $n_6$ must set its sleep time $t_{6,s}$ to $t_{6,s} - T_6^{backoff}$ (Sequence 11). Upon this adjustment, $n_6$ will wake up at time $t_6^a$ ($t_1^a < t_6^a < t_1^{send}$) in next duty cycle. The worse case for $n_6$ is that it cannot receive anything else during its active period from its neighbors such as $n_1$ and $n_5$. It will prolong its active time till receiving neighboring nodes' *beacons*, i.e. $B_1$ from $n_1$. If $B_1.Id > B_6.Id$ holds, $n_6$ will retransmit its beacon $B_6$ at time $t$ ($t = t_{n_1,a} + T_6^{backoff}$) to trigger $n_1$'s forwarding. $n_6$ switch to sleep state at $t_{6,s}$ ($t_{6,s} = t_{1,s} + T_6^{backoff}$).

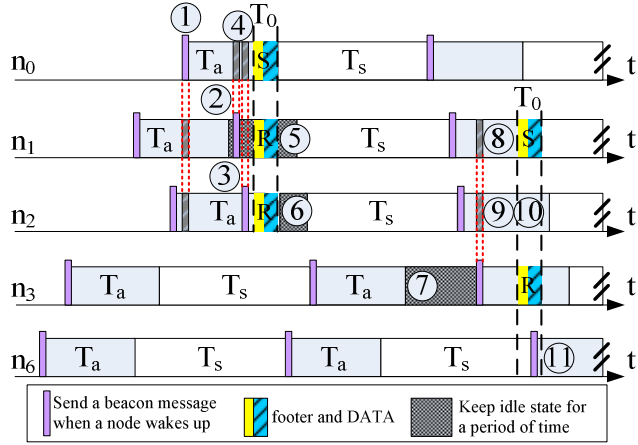

figure 5. Overview of distributed quasi-synchronized broadcast protocol of DQSB regarding the scenario in Figure 4. Every node transmits a *beacon* message when it wakes up at time $t_i^a$, indicating the maximum broadcast message ID it recently receives. Broadcast source node, such as $n_0$, makes broadcast forwarding decision according to the received *beacons*. If so, transmit at $t_i^{send} = t_i^s - T_0$, where $t_i^s$ is node $n_i$'s time to switch to sleep state. Other nodes with received *beacons* will adjust their sleep scheduling and trigger their neighbor nodes to forward based on our quasi-synchronized mechanism in Figure 3, i.e. the adjustment of $n_1$ and $n_2$ for $n_0$'s broadcast due to early sleep problem.

### 4.3. State Diagram Description

DQSB protocol has 7 states. They are idle, sleep, forward-decision, receiving, routing, transmitting and forward-unicast. Let's revisit the example shown in Figure 4. Figure 6 illustrates the state transition diagram for DQSB's running triggered by different conditions given in Table 2.

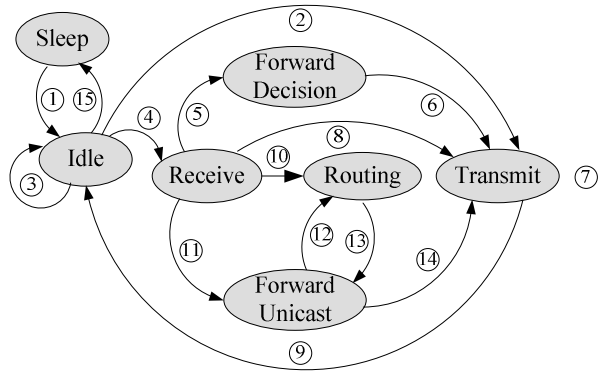

Figure 6. The state transition diagram for DQSB.

Table 2. Conditions for DQSB's state transition diagram.

| Id | Transition Condition |
|---|---|
| 1 | node $n_i$'s wakeup time $t_i^a$ arrive; |
| 2 | node $n_i$ transmit its beacon $B_i$ at time $t_i^a$; |
| 3 | node $n_i$'s active time t, $t_i^a \leq t \leq t_i^s - T_0$; |
| 4 | node $n_i$ receives *beacon*, *broadcast* or *unicast* message; |
| 5 | node $n_i$ receives *beacon*, such as $B_j$ from node $n_j$; |
| 6 | node $n_i$ satisfies $(M_i != null) \&\& (\exists m \in M_i, m.Id > B_j.Id)$; |
| 7 | node $n_i$'s forwarding broadcast time $t_i^{send} = t_i^s - T_0$; |
| 8 | node $n_i$ abort the broadcasting task at $t_i^{send}$ by received *footer*; |
| 9 | node $n_i$ finishes transmitting task and move to idle state; |
| 10 | node $n_i$ learns the next hop from received *broadcast* message; |
| 11 | node $n_i$ receives *unicast* message or *sensed data* from itself; |
| 12 | node $n_i$ looks up the route table for forwarding unicast message; |
| 13 | node $n_i$ gets the next hop for *unicast* message from route table; |
| 14 | node $n_i$ transmits the *unicast* message when the channel is idle; |
| 15 | node $n_i$'s sleep time $t_i^s$ arrive; |

1. Suppose $n_0$, as a sink, broadcasts and triggers the other nodes to transmit their sensed data to it. When *Condition 1* is established for $n_0$, $n_0$ switches to *idle* state and transmits a beacon $B_0$ with $B_0.Id = 0$ (*Condition 2*). Under *Condition 3*, $n_0$ is waiting for *beacons* from its neighbors to make forwarding decision. But $n_1$ and $n_2$ wake up before $n_0$, they can receive $n_0$'s beacon message $B_0$ and fail to receive $n_i$'s broadcast message because of early sleep problem. Both $B_1.Id$ and $B_2.Id$ are equal to -1, $B_1.Id < B_0.Id$ and $B_2.Id < B_0.Id$. They will retransmit their *beacons* before $t_0^{send}$ which will be received by $n_0$ (*Condition 4*).

2. With the received beacons $B_1$ and $B_2$, $n_0$ makes a forwarding decision (*Condition 5*). Either the value of $B_1.Id$ or $B_2.Id$ holds by *Condition 6*. Then, $n_0$ sets forwarding broadcast task at time $t_0^{send}$ and waits for transmitting (*Condition 7*). After its sending, $n_1$ and $n_2$ receive the broadcast message (*Condition 4*), and $B_1.Id$ and $B_2.Id$ are set to 0 when they wake up next time to transmit their *beacons*. If it is the sleep time of $t_{0,s}$ for $n_0$, $n_0$ turns into *sleep* state (*Condition 15*).

3. For $n_1$ and $n_2$, they learn the next hop to $n_0$ from received broadcast message and switch to *routing* state (*Condition 10*). In the next duty cycle, when their sensed data are delivered by the upper layer (*Condition 11*), they wake up and transmit their beacons according the routing table immediately if the channel is idle (*Condition 12, 13 and 14*). If there is no message to forward, then switch to *idle* state (*Condition 9*).

4. When $n_3$ and $n_6$ wake up, they perform in the same way as $n_1$ and $n_2$ in step (1). Here, it is important to notice both $n_1$ and $n_2$ receive beacons $B_3$ and $B_6$ from $n_3$ and $n_6$. Due to the case that $B_3.Id = -1$ and $B_6.Id = -1$, both $n_1$ and $n_2$ launch forwarding task at time $t_1^{send}$ and $t_2^{send}$, respectively. Because $t_1^{send}$ is earlier than $t_2^{send}$ according to the forwarding mechanism, it is the right time for *footer* to let $n_2$ abort its forwarding task (*Condition 8*) and further reduce broadcast times for energy efficiency.

5. After receiving the broadcast message, $n_3$ and $n_6$ execute what are described in step (3) to transmit their sensed data to $n_0$.

### 4.4. Further Discussion

**Property 1 (Validity and Reliability)** *With our assumption of intermittent connectivity, for any node $n_j$, there is at least one node of its neighborhoods that have received the broadcast message $n_j$ wants. Then $n_j$ finally receives the broadcast message from one of its neighbor $n_i$ within $\Delta T(j)$,*

$$\Delta T(j) \leq \Delta T(i) + t_i^s - t_j^a + T, \text{ where } T = T_a + T_s.$$

*Proof:* This property is proved based on the cases shown in Figure 3. Our assumption of intermittent connectivity implies case 4 that node receives no beacon and listens nothing else will never exist. But we introduce this case to help nodes recover from failure or join in the network, which can improve the robustness and adaptively for our protocol. We will prove this in Property 2. So, except case 4, we will deduce the latency $\Delta T_1$, $\Delta T_2$ and $\Delta T_3$ for case 1, case 2 and case 3 in Figure 3, respectively. According to the assumption, we can infer a significant result that every new broadcast message can arrive at one of $n_j$'s neighbor nodes such as node $n_i$, then this can make sure $n_j$'s successful receiving for the broadcast message from $n_i$, even if $n_j$ cannot receive the broadcast messages during time $t$, $t_{j,a} < t < t_{i,send} - T_j^{backoff}$. Because $B_j.Id$ still satisfies inequality $B_j.Id < B_i.Id$ which will trigger $n_i$'s forwarding decision. For case 1, the maximum latency of node $n_j$'s receiving for the broadcast message equals to $\Delta T^1(j)$, $\Delta T^1(j) = \Delta T(i) + t_i^s - t_j^{a,1}$. Similar to case 1, the maximum latency of case 2 satisfies the equation that $\Delta T^2(j) = \Delta T(i) + t_i^s - t_j^{a,2}$, and $\Delta T^1(j) < \Delta T^2(j)$ because of $t_j^{a,1} > t_j^{a,2}$. Case 3 is different from case 1 and case 2 for its late wake-up feature mentioned in Task 4. Obviously, this case will lead to longer latency than that can be obtained by the expression that $\Delta T^3(j) = \Delta T(i) + t_i^s - t_j^{a,3} + T$. Since $\Delta T^3(j) > \Delta T^2(j) > \Delta T^1(j)$, the maximum latency of node $n_j$'s receiving the broadcast message $\Delta T(j)$ satisfies the inequation that $\Delta T(j) \leq \Delta T(i) + t_i^s - t_j^a + T$, then Property 1 holds.

**Property 2 (Robustness and Adaptively)** *When $n_j$ recovers from failure or newly joins in a network, there is at least one node of its neighborhoods that have received the broadcast message $n_j$ wants. Then $n_j$ finally receives the broadcast message from one of its neighbor $n_i$ within $\Delta T(j)$,*

$$\Delta T(j) \leq \Delta T(i) + T, \text{ where } T = T_a + T_s.$$

*Proof:* To make DQSB be tolerant with node's failure in realistic environment and be adaptive to new node's joining, we consider the case 4 shown in Figure 3 that $n_j$ receives no *beacon* and nothing else during its active period of $[(t_j^a, t_j^s - T_0)]$. Let $\Delta T^4(j)$ be receiving latency of the broadcast message in case 4. Then, $\Delta T^4(j) = \Delta T(i) + t_i^a - t_j^a + T_a$. In the most extreme, the maximum latency occurs when node $n_j$'s wakeup time $t_j^a$ happen in node $n_i$'s sleep time $t_i^s$, and node $n_i$'s next wakeup time $t_i^a$ satisfies the condition that $t_i^a - t_j^a = T_s$. This can make the equation of $\Delta T^4(j)$ convert to $\Delta T^4(j) = \Delta T(i) + T_s + T_a$, then Property 2 has been proved upon the condition $T = T_s + T_a$.

With Property 1 and Property 2, the lower bound of latency for any node is given.

### 4.4. Less Latency Routing

We observe that a routing protocol to a sink node is related to a broadcast protocol. If a broadcast protocol constructs a good bottom-up tree path to all the other nodes in a WSN, the tree path in the reverse direction, then is a satisfying road map for all the nodes transmitting their data to the sink node. Without time synchronization in asynchronous duty-cycled WSNs, unicast protocols, such as B-MAC, WiseMAC and X-MAC, have more or less waiting latency because of the unawareness of sleep schedules of neighboring nodes. DQSB's quasi-synchronization helps the nodes to properly adjust others' wake-up schedules. For instance, $n_6$ wakes up earlier than $n_1$, and $n_1$ is earlier than $n_0$ in Figure 5. Although they do not wake up at the same time, if $n_6$ and $n_1$ adjust their sleep schedules according to quasi-synchronization mechanism, the broadcast messages sent by $n_0$ are received by all its neighbor nodes in one cycle, such as $n_1$ and $n_2$. The result of their sleep schedule relationship comes out that $n_0$ wakes up earlier than $n_1$, and $n_1$ is earlier than $n_6$. So when $n_6$ and $n_1$ wake up respectively, there is no extra waiting time for them to transmit their sensed data to the sink node $n_0$. Therefore, the less latency route for each sensed data from source to data collection node is constructed, as shown in Figure 7. The broadcast forwarding for each broadcast message only needs four nodes, i.e., $n_0$, $n_1$, $n_3$ and $n_7$. This is of great significance for energy efficient in terms of transmitting times, rather than every node's forwarding. In addition, the less latency routes are learned during broadcasting, such as $n_6 \to n_1 \to n_0$, $n_5 \to n_3 \to n_1 \to n_0$, $n_9 \to n_7 \to n_3 \to n_1 \to n_0$ and so on.

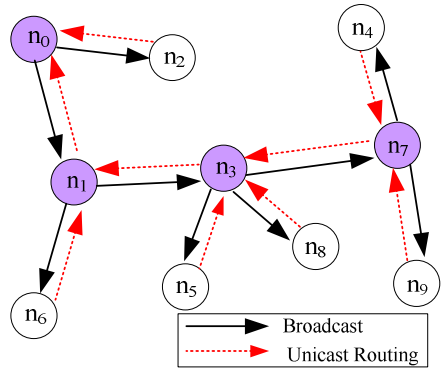

Figure 7. Every node's less latency routing path to the broadcast source node with the distributed quasi-synchronized broadcast in our DQSB protocol for the simple network scenario shown in Figure 4.

## 6. SIMULATION AND ANALYSIS

### 6.1. Simulation Setting

The ONE simulator is an open source tool for Delay-tolerant Networking (DTN), specifically designed for evaluating routing and application protocols in intermittent connective networks, such as asynchronous duty-cycled WSNs. We develop our simulator based on the ONE simulator [11] to evaluate our DQSB protocol. To satisfy the evaluation requirements, we develop extensive simulator functions based on the ONE simulator as shown in Figure 8. *Broadcast Message Event Generator* is used to generate broadcast messages in the given interval; *Random Sleep Scheduling Generator* lets randomly deployed nodes work in asynchronous duty-cycled WSNs; *Reliable Broadcast* provides distributed quasi-synchronized broadcast; and *DQSB* is applied to data collection based on *Reliable Broadcast*. We also implement Hybrid-cast and OppFlooding protocols in order to compare with our DQSB protocol.

International Journal of Wireless & Mobile Networks (IJWMN) Vol. 4, No. 3, June 2012

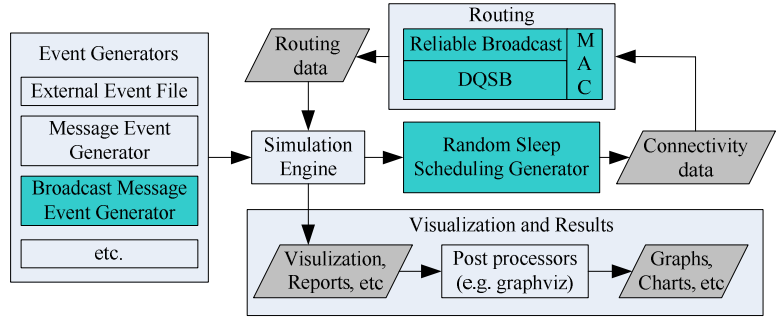

Figure 8. An Extended Simulator based on The ONE Simulator

## 6.2. Regarding Duty Cycles

We evaluate the performance in asynchronous duty-cycled WSNs with various duty cycles. In this simulation, wireless loss rate is set to 0.1, wireless communication range to 15m and transmitting speed to 250kbps. The size of a broadcast message is fixed as 512 bytes and its transmitting time of $T_0$ is 50ms. We randomly generate 10 topologies with 200 nodes, and run on each topology for 10 times.

Figure 9 illustrates the performance of forwarding times and broadcast latency, respectively. From Figure 9(a), we notice that DQSB outperforms Hybrid-cast and OppFlooding. This is because node's forwarding is triggered by its receivers in DQSB. The nodes that cannot receive the beacon will adjust their sleep schedules in order to receive broadcast messages, which is different from the other two protocols. Regarding broadcast latency in Figure 9(b), DQSB behaves particularly because its latency does not decrease in spite of the increasing of duty cycle. This contributes to DQSB's mechanism that each node in one cycle either receives a broadcast message or forwards the message received in the last cycle. Broadcast latency is related to duty cycle and forwarding times. The relationship between them is that one-hop latency follows the increasing of duty cycle in that a node launches forwarding at $t_i^{send} = t_i^s - T_0$. Generally speaking, the more nodes receive a broadcast message during one forwarding, the fewer forwarding times is. Consequently, broadcast latency in DQSB is a tradeoff between duty cycle and forwarding times.

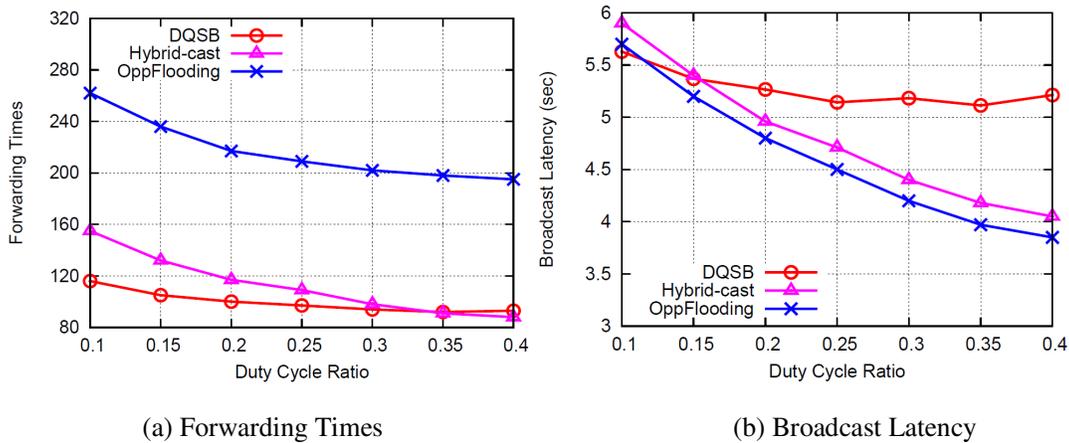

(a) Forwarding Times          (b) Broadcast Latency

Figure 9. Forwarding Times and Broadcast Latency with various duty cycles.

79

### 6.3. Regarding Network Size

Network size varies from 200 nodes to 1200 nodes and duty cycle is set to 0.2. The experiment aims to show the impact of network size in DQSB. As shown in Figure 10(a), as network size goes up, forwarding times of all the three protocols exhibit an increasing trend. However, DQSB outperforms the other two due to the same reason given before. As shown in Figure 10(b), compared with other two protocols, the broadcast latency of DQSB keeps relative tolerant and stable as the increasing in network size. So, we can conclude that DQSB can be recognized as a tradeoff between broadcast times and broadcast latency. One the one hand, node's forwarding is triggered by its receivers which helps to reduce forwarding times. On the other hand, only one of the nodes which receive beacons from neighbors forwards a broadcast message (it is an early sleep node), and the late wake-up nodes are exempt from forwarding these messages. Furthermore, quasi-synchronization greatly helps to let more nodes receive broadcast messages at each duty cycle and hence reduce forwarding times. Consequently, when network size expands, the number of nodes which can receive broadcast messages also goes up. The necessary forwarding times remains stable.

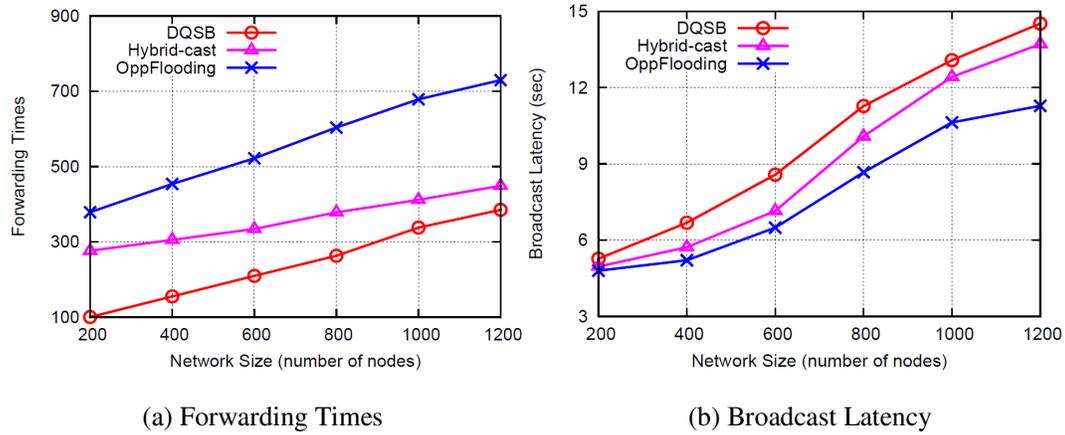

(a) Forwarding Times  (b) Broadcast Latency

Figure 10. Forwarding Times and Broadcast Latency with various network sizes.

### 6.3. Regarding Reliability with Unreliable Links

In Hybrid-cast and OppFlooding protocols, unreliable links which result in packet loss is not clearly discussed. In our DQSB protocol, *beacon* packets are sent immediately as nodes wake up. These packets are dual-folded. (1) They trigger broadcast forwarding. When receiving beacons, nodes are able to decide whether or not to forward the received broadcast messages; (2) They facilitate DQSB tolerate unreliable links. As we known, wireless links are not always reliable for many reasons in realistic environment, especially in wireless sensor networks. Consider a scenario that some nodes applying DQSB do not receive the broadcast message when a triggered node forwards a broadcast message. This does not matter because $n_i$'s failure lets itself keep the value of $B_i.Id$ in its beacon $B_i$ and it triggers its neighbors's forwarding. Thus, confronting an unreliable link, $n_i$ is able to receive a specific message if one of its neighbors receives it. The performance of reliability under unreliable links is shown in Figure 11. Regarding forwarding times, DQSB performs better than the other two protocols under the environment of unreliable links with wireless link loss rate equals to 0.1 shown in Figure 9. Even with loss rate of 0.3, broadcast latency of DQSB is still acceptable with the value of about 6 seconds.

International Journal of Wireless & Mobile Networks (IJWMN) Vol. 4, No. 3, June 2012

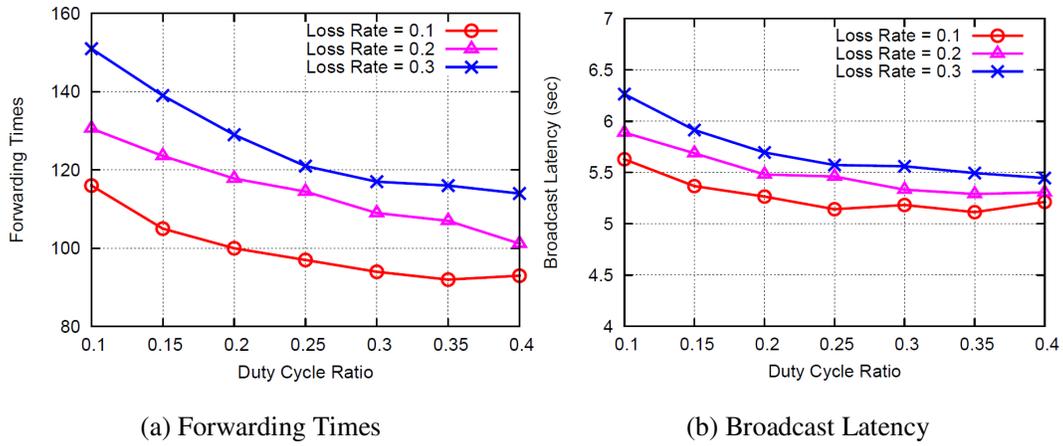

(a) Forwarding Times                (b) Broadcast Latency

Figure 11. Forwarding Times and Broadcast Latency with various link loss rates.

## 6.5. Regarding Less Latency Routing

Different from the existing multi-hop broadcast protocols in asynchronous duty-cycled WSNs, DQSB is a joint design for reliable broadcast and less latency routing paths for reverse data collection to broadcast source node, such as a sink. In this simulation, a sink node informs other nodes to send their sensed data for data collection, and sensed data are transmitted along the less latency routing paths learned by quasi-synchronized mechanism in DQSB protocol. Packet size is set to 256 bytes for sensed data and packet generating interval is [25, 35] seconds. A sink which broadcasts a message helps other nodes learn their paths from themselves to the sink when they receive the broadcast message. Each node in the network is able to complete this task because of DQSB's reliability for broadcast explained before. Figure 12 shows that DQSB behaves better in latency than LPL (Low Power Listening) which is simple and asynchronous, and adopts long preamble to make the receiver keep awake for a period of time to receive the data. So, the latency is due to the waiting time for both sender and receiver. DQSB solves this problem depending on reliable broadcast and its quasi-synchronized sleep scheduling. Figure 12 shows the less average latency for each hop in DQSB. It also illustrates that average latency for each hop in DQSB is not influenced by duty cycle due to no or less waiting time introduced.

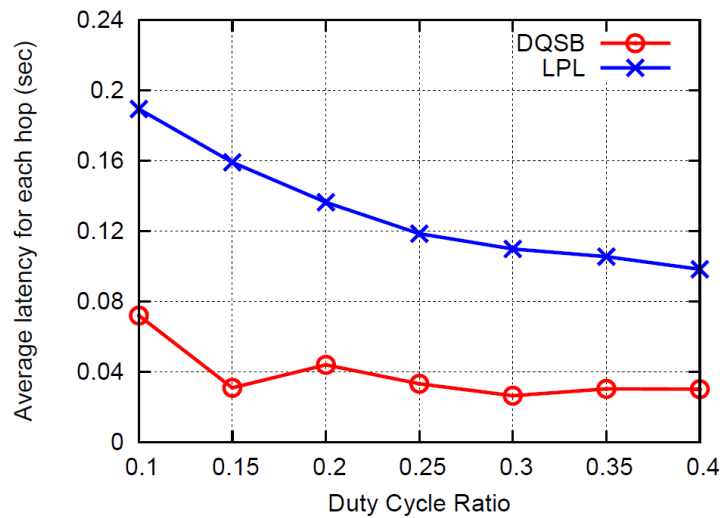

Figure 12. Average latency for each hop.

### 6.6. Selecting $\lambda$ for Intermittent Connectivity

Parameter $\lambda_0$ indicates the number of nodes that are waken at $t$. According to Poisson distribution, it is proportional to $N_R$ and inversely proportional to $T_a$. For any node $n_i$ in a network, if network density increases, $n_i$'s number of neighborhood $N_R$ increases. Thus, $n_i$'s probability to reach other nodes via its neighbor nodes may increase. Moreover, if $T_a$ increases, $n_i$'s probability to be in connection with its neighborhood increases because of the increment of node's duty cycle. But the number of neighbor nodes decreases if $\lambda_0$ declines.

We believe that if duty-cycle scheme for nodes is introduced to a connected network, the network exhibits its intermittent connection feature. We should choose the proper value of $\alpha$, and let $\lambda = \alpha \cdot \lambda_0$ in order to guarantee DQSB's feasibility, i.e., there is at least one path for any node in the network to other nodes via some intermediate nodes within a period (e.g., one cycle length T).

Table 3. Least value of $\alpha$ for $P_\lambda = 1.0$.

| Network Size | $T_a$ | $\lambda$ ($P_\lambda = 1.0$) | Least value of α |
|---|---|---|---|
| 200 nodes | 0.1 | 50 | 0.3808 |
| 200 nodes | 0.2 | 40 | 0.6093 |
| 400 nodes | 0.1 | 40 | 0.1467 |
| 400 nodes | 0.2 | 30 | 0.2201 |
| 600 nodes | 0.1 | 30 | 0.0725 |
| 600 nodes | 0.2 | 20 | 0.0966 |

In this set of simulations, we investigate the impact of the parameter $\lambda$ and select a reasonable $\alpha$ to satisfy the assumption. This implies that if this assumption holds, every node in the network receives broadcast messages due to Property 1 and Property 2 of DQSB. So, here we use broadcast success ratio $P_\lambda$ instead of network's intermittent connectivity to represent whether a network is connected. Figure 13 shows the change of $P_\lambda$ with $\lambda$ which varies from 0 to 80 in different network sizes when $T_a = 0.1$ and $T_a = 0.2$, respectively. For a given $\lambda$, $P_\lambda$ increases with network size. When $\lambda = 20$, if network size changes from 200 nodes to 600 nodes, $P_\lambda$ goes up from 0.5075 to 0.9215 in Figure 13(a) and from 0.8861 to 1.0 in Figure 13(b), respectively. Comparing Figure 13(a) with Figure 13(b), we observe that under the same $\lambda$ and network size, if $T_a$ moves larger, $P_\lambda$ also increases. When $\lambda = 20$ and network size is 200 nodes, $P_\lambda = 0.5075$ with $T_a = 0.1$ while $P_\lambda = 0.8861$ with $T_a = 0.2$. Therefore, we set $\lambda$ to $P_\lambda = 1.0$, and compute the least value of $\alpha$ in terms of the equation $\lambda = \alpha \cdot \lambda_0$, where $\lambda_0$ is available by $N_R$ and $T_a$. Least values of $\alpha$ ave given in Table 3.

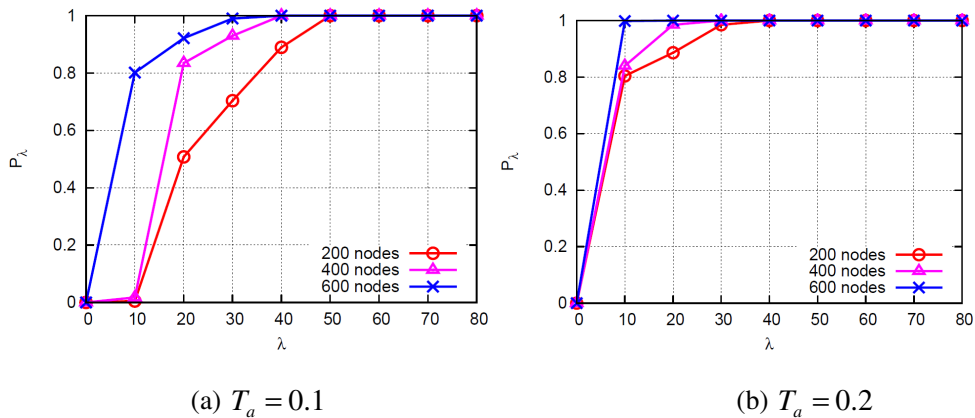

(a) $T_a = 0.1$    (b) $T_a = 0.2$

Figure 13. The impact of the parameter $\lambda$ under $T_a = 0.1$ and $T_a = 0.2$.

## 7. CONCLUSIONS

In this paper, we neither assume time synchronization which requires all neighboring nodes wake up at the same time, nor assume duty-cycled awareness which makes it difficult to use in asynchronous WSNs. A reliable broadcast protocol called DQSB is proposed by a distributed and quasi-synchronized manner for duty-cycled WSNs. Quasi-synchronization is reached after nodes execute DQSB in a local and distributed way. Under DQSB, a sink periodically broadcast. After receiving the broadcast messages, other nodes can build their paths to the sink for transmitting their sensed data. Moreover, these paths exhibit less latency because of no or very little waiting time. Simulation results show that DQSB performs well in broadcast times and keep relative tolerant broadcast latency performance. DQSB can be recognized as a tradeoff between broadcast times and broadcast latency. Further, it is still feasible under unreliable links. Our future work is to focus on applying DQSB to real WSN platforms, e.g., micaz and telosb, and investigate its performance.

## ACKNOWLEDGEMENTS

This research work is partially supported by the Natural Science Foundation of China under grant No. 60973122, the 973 Program in China under grant No. 2009CB320705, and 863 Hi-Tech Program in China under grant No. 2011AA040502. We thank all of the anonymous reviewers of this work for their valuable comments.

**Authors**

**Yun Wang** received her PhD degree from Southeast University, China in 1997 and subsequently jointed Southeast University. She is currently a full professor and vice director of the School of Computer Science and Engineering and the Key Lab of Computer Network and Information, Ministry of Education, China. From 1999 to 2000, she completed her post doctoral research in IRISA-INRIA. She then worked as a senior researcher at the University of Texas at Dallas from 2000 to 2001. She also visited Florida Atlantic University as a research fellow from 2007 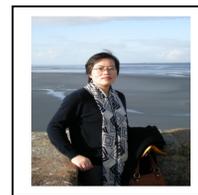 to 2008. Her research interests include distributed computing, fault tolerance and wireless networks. She has published more than 100 refereed journal and conference papers including TOC, INFOCOM, ICDCS. She is a member of the IEEE.

**Peizhong Shi** received his BS and MS degree from Zhengzhou University of Light Industry, Zhengzhou, China, in 2006 and 2009, respectively. He is currently working toward a PhD degree in the School of Computer Science and Engineering in Southeast University, Nanjing, China. His current research interests include wireless sensor networks, cross-layer design, and distributed computing. 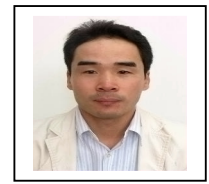

**Kai Li** received his BS degree from ShanDong University, China in 2001 and the MS and PhD degree from Southeast University, China in 2004 and 2011, respectively. He is currently a lecturer in the School of Computer Science and Engineering, Southeast University. His research interests are in distributed computing, localization in sensor networks, and p2p computing. 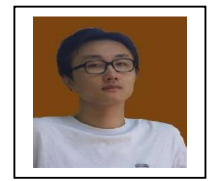

**Jie Wu** is a professor and the chairman at the Department of Computer and Information Sciences, Temple University. He has published more than 200 papers in various journal and conference proceedings. His research interests include the area of mobile computing, routing protocols, fault-tolerant computing, and interconnectionnetworks. He is a fellow of the IEEE. 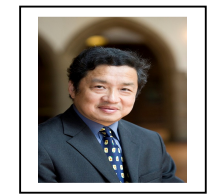